\documentclass[reprint, showpacs, amsmath,amsfonts,prb]{revtex4-1}
\usepackage{graphicx}
\usepackage{amsmath}
  \usepackage{amssymb}
  \usepackage{bm}
\usepackage[cp1251]{inputenc}
\usepackage[T1,T2A]{fontenc}

\usepackage{color}

\begin{document}
  \title{Diffraction anomalies in hybrid structures
based on\\ chalcogenide-coated opal photonic crystals }
  \author{M. M. Voronov}
    \email{mikle.voronov@coherent.ioffe.ru}
    \affiliation{Ioffe Physical-Technical Institute of the Russian Academy of Sciences,
  St. Petersburg 194021, Russia}
\author{ A. B. Pevtsov}
  \affiliation{Ioffe Physical-Technical Institute of the Russian Academy of Sciences,
 St. Petersburg 194021, Russia}
   \author{S. A. Yakovlev}
      \author{D. A. Kurdyukov}
  \affiliation{Ioffe Physical-Technical Institute of the Russian Academy of Sciences,
  St. Petersburg 194021, Russia}
    \author{V. G. Golubev}
  \affiliation{Ioffe Physical-Technical Institute of the Russian Academy of Sciences,
  St. Petersburg 194021, Russia}

\begin{abstract}
The results of  spectroscopic studies of the diffraction anomalies
(the so-called resonant Wood anomalies) in spatially-periodic hybrid
structures based on chalcogenide ($\rm{Ge_2Sb_2Te_5}$)-coated opal films of
various thickness are presented. A theoretical
analysis of spectral-angular dependencies of the Wood anomalies
has been made by means of a phenomenological approach using the concept
of the effective refractive index of waveguiding surface layer.
\end{abstract}

 \maketitle

\section{Introduction}

In recent years, investigations of periodic hybrid structures, which
are 2D and 3D photonic crystals covered with a thin metal or
dielectric film, have attracted an increasing interest.\cite{
Landstrom, Noda, Haglund,  Eggleton, Romanov} The aim of the present
work is to study character of the optical response of hybrid
structures consisting of the spatially-ordered opal film coated with
a high refractive index material. The chalcogenide compound
$\rm{Ge_2Sb_2Te_5}$ (denoted by GST225) with a refractive index
greater than 4 has been chosen to use.\cite{Raoux} Another reason
for the choice is a rather simple technique for fabrication of
$\rm{Ge_2Sb_2Te_5}$ films on the opal surface.\cite {Yakovlev} It
should be mentioned that Ge–Sb–Te based chalcogenides are used in
rewritable optical data storage applications and non-volatile phase
change memory devices. \cite{Raoux,  Burr, Karpov} In Refs.
\onlinecite{Yakovlev, Pevtsov} it is shown that the light reflection
spectra from opal$/$$\rm{Ge_2Sb_2Te_5}$ hybrid structures
demonstrate the resonant Wood anomalies, caused by the interaction
of the incident electromagnetic radiation with the hexagonally
arranged surface layer of the hybrid structure. A theoretical study
of optical spectra of these systems, like any such systems, requires
cumbersome numerical calculations and can be made analytically with
the help of the scattering matrix formalism, see, e.g., Ref.
\onlinecite{Tikhodeev}. In this work we give a simple expression
determining the light wavelengths corresponding to the spectral
positions of the Wood anomaly maxima depending on the parameters of
the hybrid structure and apply the expression to the present
experimental conditions.

\section{Phenomenological description and experiment}

The samples of the structures being studied are opal films (7-8
monolayers of a-SiO$_2$ spheres of diameter 640 nm) grown on fused
silica substrates.\cite{Pevtsov_V, Trofimova, Rybin} The
$\rm{GST225}$ layers of thickness 25-150 nm were deposited on the
surface of opal films in vacuum by the thermal deposition technique.
\cite{Yakovlev} The schematic picture of an opal$/$$\rm{GST225}$
hybrid structure and essential geometry of the experiment are shown
in Fig. 1. The light of an incandescent lamp falls on the hybrid
structure at an angle $\theta_0$ to the normal and is detected at an
angle $\theta$ (in our experiments $\theta=\theta_0$). The incidence
plane (YZ) is perpendicular to the growth plane (111) of opal film.

We shall characterize a waveguide layer consisting of a top
monolayer of opal film covered with $\rm{GST225}$ film and air by an
effective refractive index $n_{eff}$, which can be set constant in a
sufficiently wide wavelength range. It is assumed that $n_{eff}$
is the same for a direction in the plane (XY) and that
perpendicular to the plane, thus being a characteristic of
quasi-guided (or leaky) mode. \cite{Tikhodeev} The value of $n_{eff}$
depends on the direction given by horizontal projection (${\bf
k}_{\parallel}$) of the wave vector (${\bf k}$) of the wave outgoing
from the structure with respect to the crystallographic direction
$\Gamma$-M (or $\Gamma$-K), with the azimuthal angle between the
directions given by $\beta$ (see Fig.1).
\begin{figure}[h]
 \begin{center}
  \includegraphics[width=0.4\textwidth]{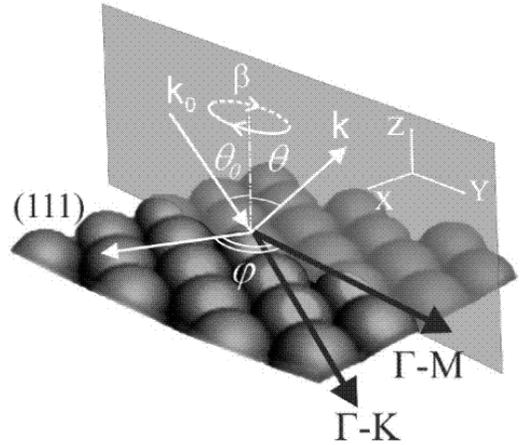}
 \end{center}
\caption{Schematic image of the surface of an
opal$/$GST225 hybrid structure and geometry of the experiment. Here
are shown the incident and outgoing wave vectors ${\bf k_0}$, ${\bf
k}$ and angles, $\theta_0$, $\theta$, azimuthal angle $\beta$ and the
main crystallographic directions $\Gamma$-M and $\Gamma$-K. The
angle $\varphi$ is that between the direction $\Gamma$-M (or
$\Gamma$-K) and some reciprocal lattice vector ${\bf G}$ (white
arrow in XY-plane). }
\end{figure}

Based on the usual diffraction condition (the Laue condition) the wave
vectors of the propagating waves can be written as ${\bf k'}={\bf
k}+{\bf G}$. Here ${\bf G}=m_1{\bf b_1}+m_2{\bf b_2}$ ($m_1$, $m_2$
are integers) are 2D reciprocal lattice vectors, where ${\bf b_1}$
and ${\bf b_2}$ are the elementary vectors (for hexagonal lattice
$|{\bf b_1}| = |{\bf b_2}| = 4\pi/\sqrt{3}d$). We proceed under the
assumption that $k'=|{\bf k'}|=\omega n_{eff}/c$ and the wave vector
${\bf k}$ of the scattered wave is a sum ${\bf k} = {\bf
k_{\parallel}}+{\bf k}_z$, where ${\bf k}_z$ is the wave vector
${\bf k}$ projection in the perpendicular direction to the waveguide
layer.  Due to the anisotropy of the system, both values $k_z$ and
$n_{eff}$ depend on $\beta$ and ${\bf G}$. The concept of the effective
refractive index for description of optical spectra of similar
systems was earlier used in Ref. \onlinecite{Landstrom}, but in that
study the parameter $k_z$ was ignored. Since $k_z$ is of the same order
of magnitude as ${\bf k}$ and ${\bf G}$, it should be taken into account.
Because of the conservation of the horizontal component ${\bf k_{\parallel}}$ we take
$k_{\parallel}=k\sin\theta$ and after simple mathematical
transformations come to the following expression for the spectral
position of the Wood anomaly peak, $\lambda_{WP}$
\begin{widetext}
\begin{equation}
\lambda_{WP}=\frac{2\pi}{G}\frac{{\sqrt{(\sin{\theta}\cos{\varphi})^{2}+(n^{2}_{eff}
-\sin^{2}{\theta})(1+a)}}-\sin\theta\cos\varphi}{1+a}
\end{equation}
\end{widetext}
Here $\varphi$ is an angle measured from $\Gamma$-M direction in the
XY plane (see Fig. 1), which defines any possible direction of vector
${\bf G}$, and $a=(k_z/G)^2$. Formally, there exists the second root of the
quadratic equation for $\lambda_{WP}$ (the negative sign before square root),
however, from a physical sense since inequality $n>\sin\theta$
is always satisfied this solution gives negative values for
$\lambda_{WP}$ and therefore should be omitted. Eq. (1)
can be used explicitly if the light scattering mainly occurs with
a single scattering vector ${\bf G}$. This would be especially well
realized for some systems with a periodic optical dielectric
constant modulation in one dimension. Nevertheless, we can also use
the above analytical expression for the systems with 2D dielectric
constant modulation like those considered here because in case of
p-polarization realized in the present experiments the
electric field distribution in the waveguide layer with a change of
angle $\theta$ mostly depends on the vectors ${\bf G}$ corresponding
to $\Gamma$-M direction. It follows from Eq. (1) that the diffraction
anomalies can be formed only at wavelengths less than the value given
by $\lambda_{WPm}=2\pi(1+n_m)/G_0$, where $G_0$ is the minimum
magnitude of the vector ${\bf G}$ and $n_m$ is the maximum refractive
index in the system.
\begin{figure}[b]
 \begin{center}
  \includegraphics[width=0.33\textwidth]{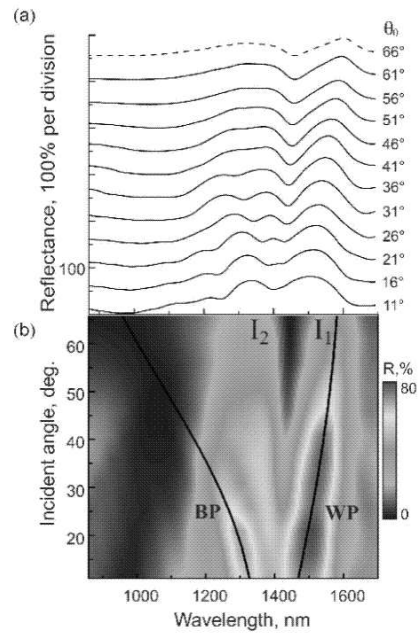}
 \end{center}
\caption{(a) Experimental reflection spectra of an
opal$/$GST225 hybrid structure for different incident angles
$\theta_0$ at $\beta = 0$. The film thickness is about 150 nm. The
spectra are shifted vertically for clarity. (b) The color map is
shown for improving visualization of the spectra. The solid lines
passing through the spectral maxima are guides for eye and
correspond to the standard Bragg diffraction (BP) from 3D opal
structure and to the Wood anomalies (WP). }
\end{figure}

Experimental reflection spectra from the opal$/$GST225 hybrid structure
with GST225-film of 150 nm thickness are shown in Fig. 2 for
different angles $\theta = \theta_0$ (at $\beta = 0^{\circ}$). In the
spectra measured  at an angle of $\theta_0=11^{\circ}$ one can distinguish
three particular features: i) typical wavelength area of Bragg
reflection in the range 1300-1350 nm, which appears due to the light
diffraction from the 3D system of planes (111) of the opal film and
shifts to shorter wavelengths when increasing the incidence angle of
light according to the Bragg formula
($\lambda_{BP}=2d_{111}\sqrt{\langle\varepsilon\rangle-\sin^2\theta_{0}}$, where
$\lambda_{BP}$ is the spectral position of 3D Bragg reflection peak,
$d_{111}$ is the period of opal structure in [111] direction,
$\langle\varepsilon\rangle$ is the average value of the dielectric constant of
opal film); ii) wide bands in the ranges 1400-1600 nm ($\rm I_1$) and
1200-1400 nm ($\rm I_2$), with reflection maxima shifting to longer
wavelengths ($\rm I_1$) or staying almost unmoved ($\rm I_2$) with
increasing the angle of light incidence (see Fig. 2b); iii) some weak
intensity peaks initiated by the Fabry-Perot interference over the whole
thickness of the hybrid structure. In the present study we focus on
case (ii).

\begin{figure}[t]
 \begin{center}
  \includegraphics[width=0.4\textwidth]{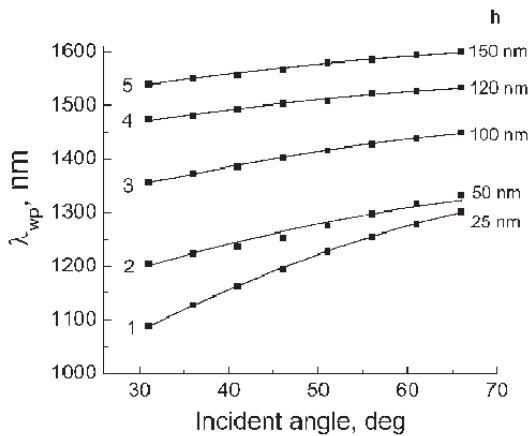}
 \end{center}
\caption{Experimental values (black squares) with corresponding
numerical fits (solid lines) made by Eq. (1) for wavelengths
$\lambda_{WP}$ for five hybrid structures differing only by the
thickness ($h$) of the GST225 film. The corresponding parameters $h$,
$n_{eff}$ and $a$ are as follows: curve 1 ($h$=25 nm, $n$=1.63,
$a$=0.12); curve 2 ($h$=50 nm,  $n$=2.31, $a$=0.56); curve 3
($h$=100 nm, $n$=3.08, $a$=0.97); curve 4 ($h$=120 nm, $n$=4.19,
$a$=1.87; curve 5 ($h$=150 nm, $n$=4.4, $a$=1.9);
$\varphi=180^{\circ}$ for all curves.}
\end{figure}

Figure 3 depicts the spectral positions (black squares) of the Wood
anomaly maxima $\lambda_{WP}$ depending on the light incidence angle
for opal$/$GST225 hybrid structures with different thickness ($h$)
of GST225 layer. The values of $h$ were determined from the SEM
micrograph of the opal/GST225 structure cleavages. It can be seen
from the figure that when thickening the film the lines
$\lambda_{WP}(\theta_0$) move to the long-wavelength region and
dependence $\lambda_{WP}(\theta_0$) is being weakened (which is
related to a decrease in the derivative
$\partial\lambda_{WP}/\partial\theta>0$); it will be explained in
the next section.
\begin{figure}[b]
 \begin{center}
  \includegraphics[width=0.4\textwidth]{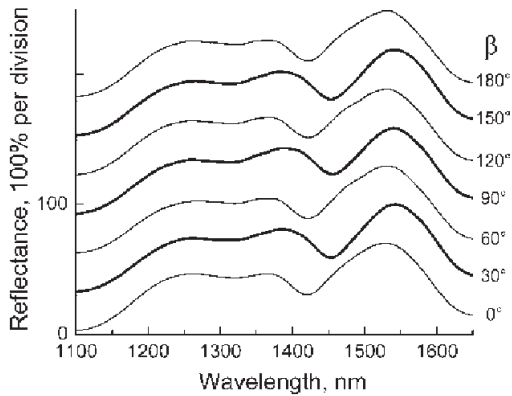}
 \end{center}
\caption{Experimental reflection spectra for the
opal/GST225 hybrid structure with GST225 film of the thickness 150 nm for different
azimuthal angles $\beta$ at an angle $\theta_0=30^{\circ}$. The spectra are shifted
vertically for clarity.}
\end{figure}

Experimental reflection spectra for different azimuthal angles
$\beta$ at the outgoing angle $\theta$ taken equal to the incidence
angle $\theta_0=30^{\circ}$ are shown in Fig. 4. Since the system under
consideration has symmetry group $C_{6v}$, under rotation of the
incidence plane at the angle $\beta=60^{\circ}$ (angle between the axes of
symmetry) the spectra must be repeated, which is indeed seen from
Fig. 4. Due to the symmetry, for a fixed value of the angle $\theta$ the
dependence $\lambda_{WP}(\beta)$ should reach extreme values
(alternating minima and maxima) at values of $\beta$ equal to the
multiples of $30^{\circ}$ that also may be seen from the figure. In the
vicinity of 1530 nm the shift of the spectrum reaches about 15
nm when $\beta$ is varying from 0 to $30^{\circ}$, a reason for it will be
discussed later.

\section{Discussion of  results}

We use Eq. (1) to describe the dependence
$\lambda_{WP}(\theta=\theta_0)$. The approximate values of
$n_{eff}\lesssim n_{GST225}$, where $n_{GST225}$ is the refractive
index of GST225, and the wavelengths about $1 {\rm \mu m}$
correspond to the vectors ${\bf G}$ of magnitude $|{\bf G}|=G_0=
4\pi/\sqrt{3}d$, where $d$ ($\approx 640$ nm) is the period of the
hexagonally-ordered surface-relief grating of the hybrid structure.
For the case of the scattering at angle $\varphi=180^{\circ}$
experimental data points in Fig. 2(b) lie well on the curves (solid
lines in Fig. 3) described by Eq. (1) under constant parameters
$n_{eff}$ and $k_z$, whose values are indicated in the caption of
the figure. The increase in $n_{eff}$ with chalcogenide film
thickness can be explained by the fact that at the same time the
average dielectric constant calculated in the effective medium
approximation increases as well. Constant values of $k_z$ for every
structure for different angles $\theta$ correspond to the situations
where the quasi-guided modes are formed. As is known, the usual
guided modes in a three--layer planar dielectric waveguide are also
characterized by special values of $k_z$. But compared to this case
where the total internal reflection condition is always satisfied
and the phase shift due to $k_z$-projection is a multiple of $2\pi$,
\cite{Yariv} for quasi-guided modes these conditions are not
fulfilled and $k_z$ can remain unchanged because of the periodicity
of the structure and the corresponding electromagnetic field
distribution (with the formation of modes with relatively high
values of the quality factor). Actually, there is a dispersion of
$k_z$ in the medium, which we do not take into account here (just as
the dispersion of $n_{eff}$). The propagation of electromagnetic
radiation in the waveguide layer is accompanied by the scattering of
the waves to the outside of the structure with a change in the
effective wave vector. In the simplest case the horizontal
projection of the wave vector of the scattered wave is different
from the value $k_0\sin\theta_0$ (where $k_0$ is the wave number of
the light in the outside medium) by a single reciprocal lattice
vector ${\bf G}$, in which case it is said that the incident
radiation resonantly interacts with a quasi-guided mode. If the
rescattering predominantly occurs with wave vector transfer $-{\bf
G}$ it can lead to the formation of Wood anomalies at an outgoing
angle $\theta = \theta_0$.

An analysis of Eq. (1) leads to the following answers for the cases
of increasing and decreasing $\lambda_{WP}$ with a change in
$\theta$:\\
a) $\partial\lambda_{WP}/\partial\theta>0$ if simultaneously
$\cos\varphi<0$ and
$n_{eff}>\sin\theta\sqrt{(\tan\varphi)^2(a+1)+a}$;\\
b) $\partial\lambda_{WP}/\partial\theta<0$ if simultaneously $\cos\varphi<0$ and\\
$n_{eff}<\sin\theta\sqrt{(\tan\varphi)^2(a+1)+a}$, and also if
$\cos\varphi>0$;\\
c) $\partial\lambda_{WP}/\partial\theta=0$ corresponds to\\
$\lambda_{WP}=2\pi n_{eff}/(G\sqrt{1+a})$ for any $\varphi$.\\
These results allow one to understand the specific character of the
light scattering at different outgoing angles. For the $\Gamma$-M direction
for all structures one can observe a monotonic increase of
the wavelength $\lambda_{WP}$ with increasing $\theta$. This
situation is in accordance with case (a) and corresponds
advantageously to the backward scattering, with wave vector
transfers ${\bf G}$ for which $\cos\varphi<0$ (for hexagonal lattice
possible angles $\varphi = 0,\pm 60^{\circ}, \pm 120^{\circ}$ and $180^{\circ}$). Note
that for arbitrary direction in the plane (XY) different from
$\Gamma$-M and $\Gamma$-K directions (see Fig. 1), due to the
asymmetry, Eq. (1) cannot be used because here one must take into
account at least two reciprocal lattice vectors corresponding to the
different values of $\varphi$. In these cases, however, due to the relatively
small anisotropy in the plane (XY), one can also expect the
appearance of a monotonic increasing dependence $\lambda_{WP}(\theta)$.
\begin{figure}[b]
 \begin{center}
  \includegraphics[width=0.4\textwidth]{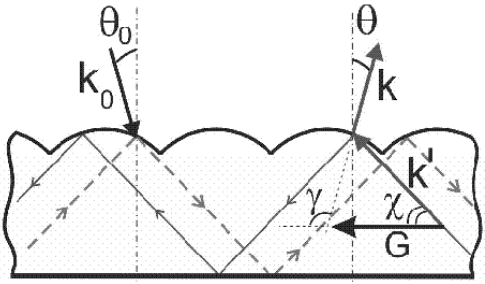}
 \end{center}
\caption{The scattering geometry: light with wave
vectors ${\bf k_0}$, ${\bf k}$ and ${\bf k'}$ propagating in a
periodic waveguide layer structure characterized by an effective
refractive index. The detailed description see in the text.}
\end{figure}

As it follows from the estimate $a=(k_z/G)^2 \sim
(\sqrt{3}nd/2\lambda)^2 \sim 1$, with an increase in the angle $\theta$
the vector ${\bf G}$ corresponding to $\varphi = 180^{\circ}$ plays an
increasingly important role because at sufficiently large values of
$\theta$ the condition in the case a) at $\varphi = 120^{\circ}$
($n_{eff}>\sin\theta\sqrt{4a+3}$) may not be satisfied, while for
the case $\varphi = 180^{\circ}$ the corresponding inequality
($n_{eff}>\sin\theta\sqrt{a}$) is weaker and consequently can be
valid (at nearly the same values of $n_{eff}$ and $a$). It is
especially typical of very thin films such as the film of thickness
25 nm. To qualitatively understand the effect of the prevailing
scattering at the angle $\theta=180^{\circ}$ (or $\theta=0$, which is not
realized in our experiments) one should consider a ray diagram (see
Fig. 5) representing the wave vectors $\bf{k}$, $\bf{k'}$ and
$\bf{G}$. With increasing $\theta$ the angle $\gamma$ between the
vectors $\bf{k}$ and $\bf{G}$ (or the angle $\chi$ between
$\bf{k'}$ and $\bf{G}$) with the opposite direction of
$\bf{k}_{\parallel}$ changes noticeably more than the angles between the
vector $\bf{k}$ (or $\bf{k'}$) and the other vectors $\bf{G}$ of the
same magnitude ($G_0$), resulting in the dependence
$\lambda_{WP}(\theta)$. In other words, with an increase in angle
$\theta$ the system behaves more and more asymmetrically, unlike the
situation where $\theta = 0$, that is due to the scattering with
wave vector transfers ${\bf G}$ of magnitude $|{{\bf G}}| =
G_0$. Note that the above backscattering may take place in the
case of 3D opal-like photonic crystals. \cite{Asher}

It may be noted that if the chalcogenide film thickness is not small
and with its change the structure remains approximately
geometrically similar to itself, then there is a quadratic dependence of
the parameter $a$ on the refractive index $n_{eff}$, that is $a \propto
k_z^2 \propto n_{eff}^2$ (in agreement with the values presented in the
caption of Fig. 3), that indirectly confirms the validity of the
proposed approach with the use of the quantities $n_{eff}$ and
$k_z$. Another point of interest mentioned above is a flattening of the
lines $\lambda_{WP}$ (i.e., decrease in the derivative
$\partial\lambda_{WP}/\partial\theta>0$) while thickening the film,
see Fig. 3. An analysis of Eq. (1) with $\varphi = 180^{\circ}$ shows
that this tendency is due to the increase in the value of parameter $a$
when increasing the film thickness $h$. A qualitative explanation of
this effect is that the thickening of the film leads to a more uniform
light scattering with wave vector transfers {\bf G} of magnitude
$|{\bf{G}}|=G_0$ and consequently to a decrease in the derivative
$\partial\lambda_{WP}/\partial\theta$.

We now address ourselves once more to Fig. 4, where the reflection spectra are
presented to demonstrate small periodic changes in $\lambda_{WP}$ in
the vicinity of 1530 nm with a change of the azimuthal angle $\beta$. To
explain it one can also exploit Eq. (1) where for the new geometry
of diffraction the vectors ${\bf G}$ of the magnitude equal to $G_0$ and
making angles $\varphi=\pm 150^{\circ}$ with $\Gamma$-K  direction should
be taken into account. Since, as it directly follows from the
effective medium approximation, the values of $n_{eff}$ and $a\propto
n_{eff}^2$ for the scattered light waves with ${\bf k_{||}}$
parallel to $\Gamma$-K direction are less than those for $\Gamma$-M
direction, it can result in about the same value of $\lambda_{WP}$
at a given angle $\theta$ (e.g., at $\theta_{0} = 30^{\circ}$, see Fig. 4).

As is seen from Fig. 2, together with the long-wavelength peaks
($\rm {I_1}$) interpreted as Wood anomalies there is a wider
short-wavelength band ($\rm {I_2}$) which can be explained as Wood
anomalies corresponding to a different quasi-guided mode; in rough
approximation their profiles can be described by the Lorentzian
functions. \cite{Fan}
\begin{figure}[h]
 \begin{center}
  \includegraphics[width=0.4\textwidth]{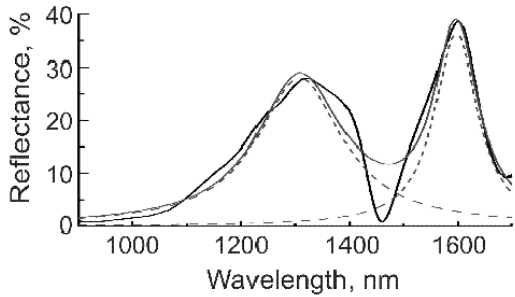}
 \end{center}
\caption{Experimental reflection spectrum (black solid line) of the
opal/GST225 hybrid structure (the film thickness $h=150$ nm) at an
angle $\theta_0 = 66^{\circ}$, and its representation by two
Lorentzians (dashed lines) and by their sum (solid line).}
\end{figure}
Figure 6 shows as an example the experimental reflection spectrum of
an opal/GST225 hybrid structure at an angle $\theta_0=66^{\circ}$ (see
Fig. 2a) decomposed by two Lorentzians. Because of the closeness of
the two peaks the Lorentz-like profiles are distorted (especially
strongly  between the maxima) even for large angles $\theta$,
therefore the damping rates $\Gamma$ of the modes are dependent on
each other. For this reason a phenomenological description of the
shapes of the spectra by using the effective values $n_{eff}$ и $k_z$ is
problematic. Another reason for this is that the value of $\Gamma$
depends in a complicated manner on the outgoing angle and for its
determination the exact numerical calculations are required. The
wider Wood anomaly peak corresponds to a greater value of $\Gamma$
and consequently to a shorter lifetime of the mode, $\tau \sim
1/\Gamma$. This mode corresponds to the situation for which
$\partial\lambda_{WP}/\partial\theta \approx 0$, that is, when the light
scattering with the different scattering vectors $\bf{G}$ (of magnitude
$|{\bf{G}}| = G_0$) occurs almost equally intensively, leading to a
smaller value of $\tau$.

\section{Conclusion}
Wood anomalies are special types of diffraction different from
Bragg one which stem from multiple wave interference and manifest themselves
in optical reflection and transmission spectra.
\cite{Hessel} Depending on the geometrical form of a system, its
dielectric and conductive properties and spectral characteristics of
electromagnetic radiation there exist two types of Wood anomalies
(Rayleigh-Wood anomaly and resonant Wood anomaly) allowing different
approximate analytical description, see, e.g., Ref.
\onlinecite{Hessel}. Diffraction anomalies of the two types, in
general, can superimpose on each other as well as on specular light
reflection from the film surface, 3D Bragg diffraction, Fabry-Perot
oscillations and the rest of the scattered radiation. The above
phenomenological description of the dependence
$\lambda_{WP}(\theta)$ can also be applied to the light transmission
spectra showing Wood anomalies to give the characteristic nearly
linear dependence $\lambda_{WP}(\theta)$, as in the case of
reflection.

In Refs. \onlinecite{Yakovlev, Henry, Gajiev, Baryshev, Moroz} the
so-called multiple Bragg diffraction in 3D photonic crystals is
considered; these systems have no additional coating serving as an
optical waveguide region. The nearly linear dependence of the
wavelength corresponding to a local maximum in reflectivity (or
transmissivity) on the angle $\theta=\theta_0$ observed
in those experiments, indicate that diffraction anomalies arising in
the Bragg reflection and transmission spectra of 3D photonic
crystals can be ascribed to appropriate effective quasi-guided modes
(characterized by values $n_{eff}$ and $k_z$). The qualitative
interpretation of the multiple Bragg diffraction in 3D opal-like
photonic crystals by involving the concept of Wood anomalies
was proposed in Ref. \onlinecite{Asher}.

It can be expected that the approach developed in the present work
can also be used to estimate the locations of Wood anomalies in the
reflection and transmission spectra of deterministic aperiodic
structures (including quasicrystals), see, e.g., Ref.
\onlinecite{Poddubny}; in this case the vector ${\bf G}$ should be
replaced by appropriate diffraction vector for quasicrystal, ${\bf
G}_{hh'}$. Despite the absence of periodicity these structures have
long-range order and therefore would provide effective quasi-guided
modes for propagation and scattering of light, leading to Wood
anomalies in the optical spectra.\cite{quasicrystals}

A further development of the proposed approach may be related to the
study of the spectral positions of the reflection maxima and minima for
waveguide grating structures characterized by a strong dispersion of
the effective refractive index $n_{eff}= n_{eff}(\omega)$ of the
waveguide layer (for example, due to plasmon resonance in metallic
films); the corresponding spectral-angular dependencies will be
obtained by solving Eq. (1) containing frequency functions
$n_{eff}(\omega)$ and $a(\omega) \propto n_{eff}^2(\omega)$.

To summarize, the experimentally measured reflection spectra fall
within a category of Wood anomalies because to explain them one has
to consider the scattering of electromagnetic radiation expressed in
terms of 2D reciprocal lattice vectors and leading to the formation
of quasi-quided modes. Thus, the developed phenomenological approach
allowed us to interpret the experimentally observed close to linear
dependence of spectral locations of the Wood anomalies on the
incident angle of light on the spatially-periodic hybrid film structure.
It can be applied to other similar systems showing diffraction
grating anomalies as well.

The authors wish to thank A.N. Poddubny for helpful discussions.
This work was supported by the scientific program №24 of the
Presidium of the Russian Academy of Sciences “Basic Foundations of
the Technology of Nanostructures and Nanomaterials” and by the
Linkage Grant of IB of BMBF at DLR.


\begin{thebibliography}{22}

\bibitem{Landstrom} L. Landstr\"om, N. Arnold, D. Brodoceanu, K. Piglmayer,
and D. B\"auerle, Appl. Phys. A {\bf 83}, 271 (2006).
\bibitem{Noda} K. Ishizaki and S. Noda, Nature \textbf{460}, 367 (2009).
\bibitem{Haglund} I. Karakurt, C. H. Adams, P. Leiderer, J. Boneberg, and
R. F. Haglund, Opt. Lett. \textbf{35}, No. 10. (2010).
\bibitem {Eggleton} B. J. Eggleton, B. Luther-Davies, and K.
Richardson, Nature Photonics \textbf{5}, 141 (2011).
\bibitem{Romanov} S. G. Romanov, A. Regensburger, A. V. Korovin, and U. Peschel,
Adv. Mater. \textbf{23}, 2515 (2011).
\bibitem{Raoux} S. Raoux, W. Welnic, and D. Ielmini, Chem. Rev. \textbf{110}, 240 (2010).
\bibitem {Yakovlev} S. A. Yakovlev, A. B. Pevtsov, P. V. Fomin, B. T. Melekh, E. Yu. Trofimova,
D. A. Kurdyukov, and V. G. Golubev, Techn. Phys. Lett. \textbf{8},
768 (2012).
\bibitem{Burr}  G. W. Burr. M. J. Breitwisch,  M. Franceschini,
D. Garetto, K. Gopalakrishnan, B. Jackson,  B. Kurdi, C. Lam,  L. A.
Lastras, A. Padilla  B. Rajendran, S. Raoux, R. S. Shenoy.
 J. Vac. Sci. Technol. B \textbf{28}, 223 (2010).
 \bibitem{Karpov}M. Nardone, M. Simon,  I. V. Karpov, V. G. Karpov.
 J. Appl. Phys. \textbf{112}, 071101 (2012).
\bibitem{Pevtsov} A.B. Pevtsov, A.N. Poddubny, S.A. Yakovlev, D.A. Kurdyukov, and V.G. Golubev,
J. Appl. Phys. \textbf{113}, 144311 (2013).
\bibitem{Tikhodeev}
S.G. Tikhodeev, A.L. Yablonskii, E.A. Muljarov, N.A. Gippius, and T.
Ishihara, Phys. Rev. B \textbf{66}, 045102 (2002).
\bibitem{Pevtsov_V} A. B. Pevtsov, S. A. Grudinkin, A. N. Poddubny, S. F. Kaplan, D. A.
Kurdyukov, and V. G. Golubev, Semiconductors 44, 1537 (2010).
\bibitem{Trofimova}E. Y. Trofimova, A. E. Aleksenskii, S. A. Grudinkin, I. V.
Korkin, D. A. Kurdyukov, and V. G. Golubev, Colloid J. \textbf{73},
546 (2011).
\bibitem{Rybin}M. V. Rybin, I. S. Sinev, A. K. Samusev, K. B. Samusev, E. Y. Trofimova, D. A. Kurdyukov,
V. G. Golubev, and M. F. Limonov, Phys. Rev. B \textbf{87}, 125131
(2013).
\bibitem{Yariv} A. Yariv and P. Yeh, Optical Waves in Crystals: Propagation and Control
of Laser Radiation (Wiley, New York, 1984).
\bibitem{Asher}
A. Tikhonov, J. Bodin, and S.A. Asher, Phys. Rev. B \textbf{80},
235125 (2009).
\bibitem {Fan} S. Fan, J. D. Joannopoulos, Phys. Rev. B \textbf{65}, 235112 (2002).
\bibitem{Hessel}
A. Hessel and A.A. Oliner, Appl. Opt. \textbf{4}, 1275 (1965).
\bibitem{Henry}
Henry M. van Driel and Willem L. Vos, Phys. Rev. B \textbf{62}, 9872 (2000).
\bibitem{Gajiev}
G. M. Gajiev, V. G. Golubev, D. A. Kurdyukov, A. V. Medvedev, A. B. Pevtsov, A. V. Sel’kin, and V. V. Travnikov, Phys. Rev. B \textbf{72}, 205115 (2005).
\bibitem{Baryshev}
A.V. Baryshev, A.B. Khanikaev, R. Fujikawa, H. Uchida, and M. Inoue,
Phys. Rev. B \textbf{76}, 014305 (2007).
\bibitem{Moroz}
A.V. Moroz, M.F. Limonov, M.V. Rybin, K.B. Samusev, Physics of the
Solid State, \textbf{53} 1105 (2011).
\bibitem{Poddubny}
A.N. Poddubny, Phys. Rev. B \textbf{83}, 075106 (2011).
\bibitem{quasicrystals}
The simplest system here is a 1D resonant grating waveguide structure with
a rectangular profile composed of alternating segments with two different
lengths to form a quasicrystal (for example, Fibonacci quasicrystal).
The study of the light reflection from such a structure can be made by
a method applicable to the case of the analogous periodic structure,
 [see D. Rosenblatt, A. Sharon, and A. A. Friesem, IEEE
 J. Quantum Electron. {\textbf {33}}, 2038 (1997)], that
leads to a conclusion about the possibility of an appearance of Wood
anomalies in the reflection spectra for some non-periodic structures
having an average period (in particular, for a weakly disordered
periodic structure).

\end{thebibliography}
\end{document}